\documentclass[aps,prb,showpacs,twocolumn]{revtex4-1}

\usepackage{graphicx}
\usepackage{amsmath}
\usepackage{amssymb}

\newcommand{\bnen}{\begin{equation}}
\newcommand{\eden}{\end{equation}}
\newcommand{\bean}{\begin{eqnarray}}
\newcommand{\eean}{\end{eqnarray}}

\newcommand{\bna}{\begin{array}}
\newcommand{\eda}{\end{array}}

\begin{document}

\title{Josephson current in ballistic superconductor-graphene systems}

\author{Imre Hagym\'asi$^1$}
\author{ Andor Korm\'anyos$^2$}
\author{ J{\'o}zsef Cserti$^1$}

\affiliation{$^1$Department of Physics of Complex Systems,
E{\"o}tv{\"o}s University, H-1117 Budapest, P\'azm\'any P{\'e}ter s{\'e}t\'any 1/A, Hungary }
\affiliation{$^2$Department of Physics, Lancaster University, Lancaster, LA1 4YB, UK}

\begin{abstract}
We calculate the phase, the temperature and the junction length dependence of the supercurrent
for ballistic graphene  Josephson-junctions. For low temperatures
we find non-sinusoidal dependence of the supercurrent on the superconductor phase difference
for both short and long junctions. The skewness, which characterizes the deviation of the current-phase
relation from a simple sinusoidal one, shows a linear dependence on the critical current
for  small currents. We discuss the similarities and
differences with respect to the classical theory of Josephson junctions, where the weak link
is formed by a diffusive or ballistic metal. The relation to other recent theoretical results on graphene
Josephson junctions is pointed out and the possible experimental relevance of our work is considered
as well.

\end{abstract}


\maketitle

\section{Introduction}
The peculiar electronic properties of graphene first observed experimentally
by Geim et al.~\cite{Novoselov_graphene-1} and Zhang et al.~\cite{Zhang_graphene:ref}
can accurately be described by massless two dimensional Dirac fermion excitations
(for reviews on the physics of graphene see,
e.g, Refs.~\onlinecite{Katsnelson:rev,Katsnelson_Novoselov:rev,Geim_Novoselov:rev,neto:109}).

Owing to the proximity effect a superconductor can induce non-zero pair-potential
in the graphene as well\cite{fagas:224510}.
Such graphene-superconductor hybrid structures and in particular the  Andreev-reflection
taking place at the graphene-superconductor interface was first studied theoretically
by Beenakker~\cite{PhysRevLett.97.067007} (for a review on Andreev reflection
in graphene see Ref.~\onlinecite{RevModPhys.80.1337}).
Soon after Beenakker's pioneering work,
supercurrent between two superconducting electrodes on top of a graphene monolayer
has been observed experimentally by Heersche \emph{et al} ~\onlinecite{Heersche_supra:ref},
and later in Refs.~\onlinecite{
Heersche-further-publ,Miao:cikk,EPL.79.57008_Shailos:cikk,du:184507,ojeda}. In particular, the
experimental results of Ref.~\onlinecite{Miao:cikk} attest to the ballistic propagation
of quasiparticles in graphene-superconductor hybrid structures, whereas the experiment of
Du \emph{et al}\cite{du:184507} gave evidence that it was possible to fabricate
transparent SG interfaces. These experiments have
also sparked considerable theoretical interest in superconductor-graphene-superconductor (SGS)
heterostructures.  The short junction limit, where the  coherence length $\xi = \hbar v_F /\Delta_0$
(here $v_F$ is the graphene Fermi velocity  and $\Delta_0$ is the superconducting gap)
is smaller than the length $L$ of the junction, was first studied by Titov and Beenakker~\cite{titov:041401}
assuming ballistic graphene. In the opposite, long junction limit the density of states of the Andreev levels
was calculated first by Titov, Ossipov and Beenakker~\cite{PhysRevB.75.045417}.
Subsequently, numerous other theoretical works investigated the Josephson current in SGS
structures~\cite{PhysRevB.74.241403,MoghaddamApplPhys:cikk,PhysRevB.76.054513,du:184507,PhysRevB.77.075409,
JPhysC.20.145218,PhysRevB.78.024504}.
The tunneling effect in SG structures has been studied in several
works~\cite{PhysRevB.77.205425,PhysRevB.79.115131,bhattacharjee:217001,PhysRevB.76.184514,PhysRevB.74.180501} as well.
Other works in the field of graphene-superconductor heterostructures include studies
 on crossed Andreev reflection in a graphene bipolar
transistor~\cite{cayssol:147001,PhysRevB.80.014513},
on $s$- and $d$-wave SG junctions~\cite{PhysRevLett.99.147001,PhysRevB.77.064507} and
on ferromagnetic SG structures~\cite{PhysRevB.78.115413,PhysRevB.78.014514}.
Very recently, using a phase-sensitive SQUID interferometry technique  Chialvo et al.\ has studied
experimentally the current-phase relation (C$\Phi$R) of graphene Josephson
junctions~\cite{Exp_Chialvo_graphene_supra:cikk}.

In this work we calculate the Josephson current in SGS structure as a function of the
superconductor phase difference, the temperature and the length of the junction.
In our theoretical treatment we adapted the method used by Brouwer and Beenakker
for metallic chaotic Josephson junctions~\cite{BrouwerChaos:cikk}.
The  approach allows to obtain results for finite temperature and is valid for
junctions of arbitrary length.
We note that this method has already been applied for calculating the persistent current
through a \textit{n-p} junction in graphene\cite{PhysRevB.77.075409}. Wherever possible, we
compare our results to previous ones derived for superconductor-normal conductor-superconductor
(SNS) junctions, where the normal conductor is a ballistic metal.

The rest of the paper is organized in the following way: in the next section we introduce the
theoretical approach that we use to obtain the Josephson current. In Section~\ref{numerics}
we present and discuss the results of numerical calculations. Finally, in Section~\ref{summary} we
give a brief summary.

\section{Theoretical concept}
\label{theory}
We consider a Josephson junction in the $x$-$y$ plane.
The normal graphene region (G) at $|x| < L/2$  separates the two superconducting regions formed
by covering the graphene layer by two superconducting electrodes (S)
in the regions $x<-L/2$ and $x > L/2$  (for the geometry see Ref.~\onlinecite{titov:041401}).
The width of the Josephson junction along the $y$ axis is $W$.
Owing to the valley degeneracy of the Hamiltonian, the full Dirac-Boguliubov--de Gennes (DBdG)
equations for graphene-superconductor systems
decouple to two four by four, reduced Hamiltonians that are related to each other by a unitary transformation
(see, e.g., Ref.~\onlinecite{PhysRevLett.97.067007}).
We now take the one corresponding to the valley ${\bf K}$.
Then the quasi particle excitations in the SGS systems are described by the reduced DBdG equations:
\begin{equation}\label{DBdG:eq}
    \left(
      \begin{array}{cc}
        H_0 - \mu & \Delta(x,y) \\[1ex]
        \Delta^*(x,y)  & \mu -H_0 \\
      \end{array}
    \right) \Psi = \varepsilon \Psi,
\end{equation}
where $H_0=-i \hbar v_F (\sigma_x \partial_x + \sigma_y \partial_y) + U(x,y)\, \sigma_0 $ is the Dirac Hamiltonian.
Here $\sigma_x$ and $\sigma_y$ are Pauli matrices, $\sigma_0$ is the unit matrix and,
$\mu$ is the chemical potential and $\varepsilon >0 $ is the excitation energy.
The superconductor electrodes are doped by the potential $U(x,y)=U_0 \Theta(|x|-L/2)$
(here $U_0 < 0$ and constant, and $\Theta(x)$ is the Heaviside function).
The wave function $\Psi = {(\Psi_e, \Psi_h)}^T$ is
comprised  of  electron $\Psi_e$ and hole  $\Psi_h$ wave functions
which have opposite spin and valley indices. For the  pair potential $\Delta(x,y)$
we assume a simple model: its magnitude $\Delta_0$ is constant in the S regions,
changes step-function-like at the SG interfaces (so called ``rigid boundary condition``,
see Ref.~\onlinecite{CPR-RevModPhys})
and is zero in the normal conducting region. Similarly,
we assume that the its phase is piecewise constant in the S regions.
Hence, the pair potential is given by
$\Delta(x,y) =  \Delta_0 \, e^{-i\phi/2}$
for $x<-L/2$ and
$\Delta(x,y) =  \Delta_0 \, e^{i\phi/2}$
for $x > L/2$.  Band bending or other effects of the superconducting electrodes are neglected
(see e.g. Ref.~\onlinecite{khomyakov} for the discussions of some of these effects in the case
of normal conducting metal electrodes).

The Josephson current at finite temperature is given by\cite{BrouwerChaos:cikk}
\begin{equation}
    I= -2 k_{\textrm{B}} T\, \frac{4e}{\hbar}\, \frac{d}{d \phi}\, \int_0^\infty \,
    d \varepsilon \varrho (\varepsilon) \ln \left[ 2 \cosh \left(\frac{\varepsilon}{2 k_{\textrm{B}} T}\right)\right],
\label{eq:josephson-curr}
\end{equation}
where $\phi$ is the phase difference across the junction,
$\varrho (\varepsilon)$  is the density of states of the Andreev levels.
The factor of 4 accounts for the spin and valley degeneracy.
As one can see from Eq.~(\ref{eq:josephson-curr}), a necessary
 step to calculate the Josephson current is to obtain the density of states of
Andreev bound states in the SGS junction. To this end one can in principle proceed in the following way:
one can write down a trial wave function in all three regions of the SGS structure.
The boundary conditions at the two graphene-superconductor
boundaries of the SGS junction then result in a secular equation $\mathcal{F}(\varepsilon)=0$
whose solutions $\varepsilon_i$ give the energies of the Andreev bound states.
For finite $U_0$ we obtained a $8 \times 8$ determinant for the secular equation
(since this determinant is quite lengthy here we do not present its detailed form).
Once the energy levels $\varepsilon_i$ of the SGS junction are known,
the density of state is given by $\varrho (\varepsilon) = \sum_i \delta(\varepsilon-\varepsilon_i)$.

However, the above outlined  method is numerically quite cumbersome since one has to search for
the zeros of the secular equation $\mathcal{F}(\varepsilon)=0$.
To overcome this problem we now follow the method used by Brouwer and Beenakker in
Ref.~\onlinecite{BrouwerChaos:cikk}.
They rewrote the expression for the Josephson current given in Eq.~(\ref{eq:josephson-curr})
in a more convenient form. Here we only summarize the main steps of the derivation.
The secular equation can be written as $\mathcal{F}(\varepsilon)=
\mathcal{F}_0 \Pi_i ((\varepsilon-\varepsilon_i))=0$, where $\mathcal{F}_0$ is a
function of $\varepsilon$ but does not have zeros in the complex plane.
Thus it is easy to see that the density of states can be expressed as
\begin{equation}\label{DOS:eq}
    \varrho (\varepsilon) = -\frac{1}{\pi}\, \frac{d }{d \varepsilon}\,
    \textrm{Im} \ln  \mathcal{F}(\varepsilon+i 0^+),
\end{equation}
where $0^+$ is a positive infinitesimal.
Using the analytic properties of $\mathcal{F}$ in the upper half of the complex $\varepsilon$-plane
and the fact that under the change $\varepsilon \to -\varepsilon $ the function $\mathcal{F}$ goes over
into its complex conjugate (physically, this follows from the electron-hole symmetry),
the $\varepsilon$-integration can be extended from $-\infty$ to $\infty$.
Finally, after performing a partial integration the Josephson current in Eq.~(\ref{eq:josephson-curr})
can be rewritten as
\begin{equation}\label{Josephson-2:eq}
     I= - \frac{2e}{i \pi \hbar}\, \frac{d}{d \phi}\, \int_{-\infty +i 0^+}^{\infty +i 0^+}\,
    d \varepsilon \, \tanh \left(\frac{\varepsilon}{2 k_{\textrm{B}} T}\right) \ln \mathcal{F}(\varepsilon).
\end{equation}
Now closing the integration contour in the upper half of the complex $\varepsilon$-plane
and applying the residue theorem the Josephson current becomes
\begin{equation}\label{Josephson-3:eq}
     I= - \frac{4 e}{\hbar}\,  2 k_{\textrm{B}} T\,  \sum_{n=0}^\infty
     \frac{d}{d \phi}\, \ln \mathcal{ F}(i \omega_n),
\end{equation}
where $i \omega_n = i(2n+1)\pi k_{\textrm{B}} T$ are Matsubara frequencies.
Note that in our model $\ln \mathcal{ F}(\varepsilon)$ has no singularities for $\textrm{Im }\varepsilon >0$,
thus the poles of the integrand in Eq.~(\ref{Josephson-2:eq}) come only from the hyperbolic tangent function.
The main advantage of this result is that one does not need to obtain explicitly the
solutions of  the secular equation $\mathcal{F}(\varepsilon)=0$.
Moreover, this method immediately gives the finite temperature dependence of the Josephson current.
In the numerical calculations it turns out that the sum in Eq.~(\ref{Josephson-3:eq}) is rapidly convergent
and usually one need to include only a finite number of terms.
A similar result has been found for the persistent current through a \textit{n-p} junction
in graphene\cite{PhysRevB.77.075409}.

We now consider the experimentally relevant case of highly doped superconductor electrodes,
ie, the limit $U_0 \to -\infty$. By matching the wave functions at the graphene-superconductor
boundaries  of the SGS structure we found the same secular equation $\mathcal{ F}(\varepsilon,q_m)=0$ as that
obtained by Titov and Beenakker using the transfer matrix method
(see Eq.~(14) in Ref.~\onlinecite{titov:041401}).
We used the `infinite mass' boundary conditions\cite{PhysRevLett.96.246802} at $y=0$ and $y=W$ for which
$q_m = (m+1/2) \pi/W$, where $m=0,1,2,\dots$ (for $W\gg L $ the choice of the boundary conditions is irrelevant).
For a given $m$ the solutions of the quantization condition $\mathcal{ F}(\varepsilon,q_m)=0$ give the
Andreev energy levels $\varepsilon_m$ for $\varepsilon_m < |\Delta|$.
The secular equation $\mathcal{ F}(\varepsilon,q_m)=0$ is valid both in short and long junction limit\cite{titov:041401}.
One can show that $\mathcal{ F}(-\varepsilon + i 0^+, q)= \mathcal{ F}^*(\varepsilon + i 0^+, q)$ which is a
necessary condition\cite{BrouwerChaos:cikk}
for writing the Josephson current in the form of Eq.~(\ref{Josephson-2:eq}).

The current contribution from each propagating mode with transverse momentum $q_m$ can be calculated
separately and the total current is the sum over these contributions:
\begin{equation}\label{Josephson-4:eq}
     I= - \frac{4 e}{\hbar}\,  2 k_{\textrm{B}} T\,\sum_{m=1}^M  \sum_{n=0}^\infty
    \frac{d}{d \phi}\, \ln \mathcal{ F}(i \omega_n, q_m),
\end{equation}
where $M$ is the number of propagating modes and the function $\mathcal{ F}(\varepsilon, q)$ determines
the energy levels for a given transverse momentum $q$.
This equation is our starting point for calculating the supercurrent through a graphene based Josephson junction.

Further analytical progress can be made in the short junction limit ($L \ll \xi$) because
the Andreev levels $\varepsilon_m$ can in that case be obtained in a closed form
(see  Eq.~(16) in Ref.~\onlinecite{titov:041401}).
Similarly, the summation over the Matsubara frequencies in Eq.~(\ref{Josephson-4:eq}) can be performed
analytically using the identity
$\sum_{k=0}^\infty 1/\left[(2 k + 1)^2 + x^2\right]= (\pi \tanh (\pi x/2)/(4 x)$
(see Ref.~\onlinecite{Grads:book}) and we find
\begin{gather}
I=\frac{e\Delta_0^2(T)}{\hbar}\, \sin\phi \sum_{m=0}^{\infty}\frac{\tau_m}{\varepsilon_m}
\tanh \left(\frac{\varepsilon_m}{2k_\textrm{B} T}\right).
\label{eq:short_aram_alt}
\end{gather}
Here $\varepsilon_m$ and $\tau_m$ can be found in Ref.~\onlinecite{titov:041401},
while the temperature dependence of the superconductor gap $\left|\Delta  \right| = \Delta_0(T)$
for $s$-wave superconductors is given by
\begin{equation}\label{gap_T:eq}
    \ln \frac{\Delta_0(0)}{\Delta_0(T)} =
    2 \sum_{n=1}^\infty \, {\left( -1 \right)}^{n+1} K_0\left(n\frac{\Delta_0(T)}{k_\textrm{B} T}\right),
\end{equation}
where $K_0(x)$ is the zero order modified Bessel function,
$\Delta_0(0) = (e^\gamma /\pi ) k_\textrm{B} T_c = 0.567 k_\textrm{B} T_c$
and $\gamma $ is the Euler's constant\cite{Abrikosov_Gorkov_Dzyaloshinksii:book}.
For zero temperature from (\ref{eq:short_aram_alt}) one can arrive at the same expression for the Josephson current
as that obtained  by Titov and Beenakker (Ref.~\onlinecite{titov:041401}).
For finite temperatures the summation over the transversal modes $m$ in Eq.~(\ref{Josephson-4:eq})
cannot be evaluated analytically but numerically can easily be treated.

\section{Numerical results}
\label{numerics}
We now present the results of numerical calculations for the Josephson current using
the most general formula given by Eq.~(\ref{Josephson-4:eq}).
\begin{figure}[!ht]
\includegraphics[scale=0.3]{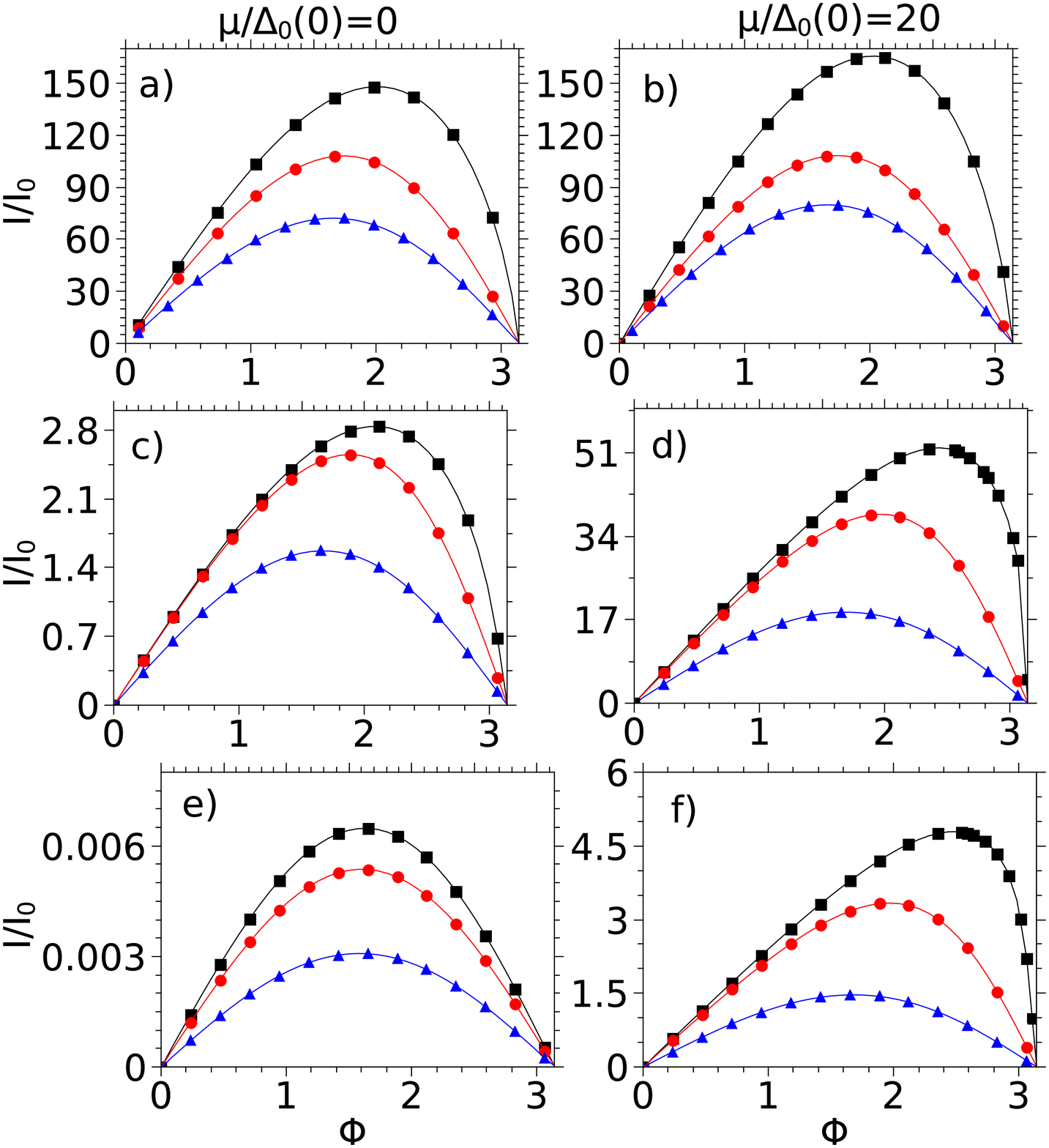}
\caption{\label{fig:cprs}
(color online)
The supercurrent (in units of $I_0= e\Delta_0(0)/\hbar$)
as a function of the phase difference $\phi$.
The parameters are as follows: In 
(a) and (b)
[short junction limit] $\xi/L=20$ and  $T/T_c = 0, 0.53, 0.71$
(black $\Box$, red $\circ$ and blue $\triangle$, respectively).
In 
(c) and (d) [here $L\simeq\xi$] we used $\xi/L=0.91$ and  $T/T_c = 0, 0.18, 0.35$
(black $\Box$, red $\circ$ and blue $\triangle$, respectively).
In 
(e) and (f) we consider long junctions [$\xi/L=0.05$].
In 
(e) $T/T_c = 0, 0.035, 0.053$ (black $\Box$, red $\circ$ and blue $\triangle$, respectively).
In (f) $T/T_c= 0, 0.018, 0.035$ (black $\Box$, red $\circ$ and blue $\triangle$, respectively).
The chemical potential is $\mu=0$ in (a), (c) and (e), whereas it is $\mu/\Delta_0=20$ in
(b), (d) and (f). The width of the junction is $\xi/W=0.05$ in all cases.
}
\end{figure}
Figure~\ref{fig:cprs} shows the supercurrent  as a function of phase $\phi$ for a number
of interesting case: for short  ($L \ll \xi$), intermediate ($L \simeq \xi$) and
long ($L \gg \xi$) junctions, assuming $\mu=0$  and for finite $\mu/\Delta_0$ as well.
One can make the following general observations: a) the maximum current increases by increasing
the doping ($\mu/\Delta_0$ value) and by decreasing the temperature or the junction length;
b) at higher temperatures the current shows a  simple sinusoidal dependence on the phase
in all of the cases while at low temperatures the position of the maxima of the
currents are shifted to the right resulting in a skewness of the curves.
Following Ref.~\onlinecite{Exp_Chialvo_graphene_supra:cikk}
we define the  skewness  by $S=2\phi_{\textrm{max}}/\pi -1$, where $\phi_{\textrm{max}}$
is the position of the maxima of the supercurrent.
Both the tendency to simple harmonic dependence for $T\rightarrow T_c $ and
a positive skewness (i.e. $\phi_{\textrm{max}} > \pi/2$) are in line with  previous
calculations on Josephson current in weak links which comprise a normal conducting
metal or a tunnel barrier  and assume that the pair potential changes
abruptly at the normal-superconductor interface (see e.g. Ref.~\onlinecite{CPR-RevModPhys} for a
review).

We find especially interesting the results shown in Figs.~\ref{fig:cprs}(e) and (f).
In the long junction limit for $\mu=0$ we find that the skewness is very small even at $T/T_c=0$
[the curve denoted by black squares in Fig.~\ref{fig:cprs}(e)] thus the C$\Phi$R   
resembles a harmonic dependence. In contrast, still in the  long junction limit but
for $\mu/\Delta_0=20$ and $T/T_c=0 0$ [black squares in Fig.~\ref{fig:cprs}(f)] we see
that the current depends
almost linearly on the phase and the curve resembles a saw-tooth.
(The transition to a saw-tooth-like dependence can already be seen in Fig.~\ref{fig:cprs}(d) where
 $L/\xi=1.1$, c.f. Fig.~\ref{fig:cprs}(b) showing the short junction limit.)
It is interesting to note  that the theoretical result for clean, long SNS junctions at low temperature is
a saw-toothed C$\Phi$R 
\cite{ishii,CPR-RevModPhys}. Our numerics suggests  that
for SGS junctions in the same limit the saw-tooth is somewhat rounded-off and
the slope of the curve is finite when $\phi\rightarrow\pi$. Thus, the C$\Phi$R in long, clean
SGS junction seems to be closely resembling of, but not identical to the corresponding case in
SNS junctions.

\begin{figure}[!ht]
\includegraphics[scale=0.55]{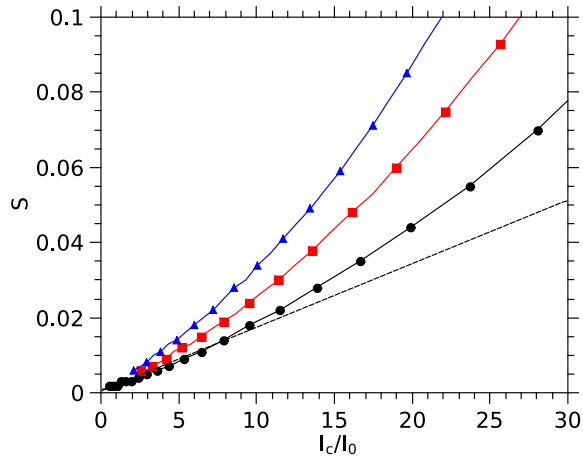}
\caption{\label{skewness_A:fig} (color online)
The skewness $S$ as a function of the critical current $I_c$ 
for different coherence lengths $\xi$.
The parameters are $\xi/L=0.35$, $\xi/W=0.0077$ (black dots),
$\xi/L=1.05$, $\xi/W=0.0231$ (red squares) and $\xi/L=1.75$, $\xi/W=0.0385$ (blue triangles).
The lines are guides to the eye.
The ratio of the chemical potential and the superconducting gap was $\mu/\Delta_0=10$.
The dotted line shows that for small critical currents the skewness depends linearly on $I_c$.}
\end{figure}
We calculated the skewness $S$ as a function the critical supercurrent $I_c$ (the value of the
current at $\phi_{\textrm{max}}$) and plotted the results for three different $\xi$ values, while
keeping the junction length $L$ and width $W$ constant.
The used $\xi/L$ values go from $\xi/L=0.35$ (long junction limit)
to $\xi/L=1.75$ (short junction limit).
There are two  important things to notice in Fig.~\ref{skewness_A:fig}
for small critical currents: a) for a given junction length $L$,
as $I_c\rightarrow 0$ (for higher
temperatures) the skewness $S$ also goes to zero, i.e. the C$\Phi$R is
approaching a simple sinusoidal form; b)  $S$ depends linearly on $I_c$ for small critical currents,
while at  larger $I_c$ the dependence clearly deviates from a simple linear relation.
The skewness has recently been measured in Ref.\onlinecite{Exp_Chialvo_graphene_supra:cikk} and the
case of  $\xi/L=0.35$, $\xi/W=0.0077$, shown by black dots in Fig.~\ref{skewness_A:fig},
in principle  corresponds to that  of sample B in this experiment
(with estimated coherence length of  $\xi \approx 100$~nm~\cite{private:note}).
Our calculations give a smaller slope than the measurements  in Ref.~\onlinecite{Exp_Chialvo_graphene_supra:cikk}.
According to our numerics, the  larger slopes observed in this experiment would be attainable in
the short junction limit.
 Note however, that the exact slope would also depend on the value of the chemical potential which
was not known, and importantly, the samples in the experiment
of Ref.~\onlinecite{Exp_Chialvo_graphene_supra:cikk} are likely to have been in the
quasi-diffusive limit, therefore we cannot expect  quantitative agreement with our ballistic theory.

\begin{figure}[!ht]
\includegraphics[scale=0.4]{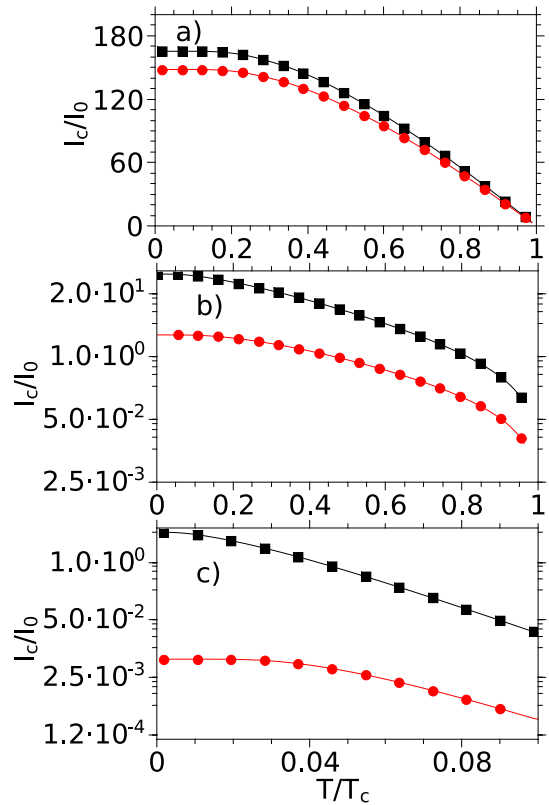}
\caption{\label{kT_depend:fig}
(color online)
The critical current $I_c$ 
as a function of $T/T_c$.
The parameters are  $\xi/L=20$ in (a), $\xi/L=0.91$ in (b) and $\xi/L=0.05$ in (c).
In (b) and (c) we used logarithmic scale.
Red $\circ$ denote the results for $\mu=0$ and black $\Box$ for $\mu/\Delta_0(0)=20$.
The width of the sample was fixed : $\xi/W=0.05$.
}
\end{figure}

We have also calculated the temperature dependence of the critical current for short, intermediate
and long junctions, taking two values of the chemical potential $\mu$.
The temperature dependence of the pair potential was taken into account using Eq.~(\ref{gap_T:eq}).
The results are shown in Fig.~\ref{kT_depend:fig}.
At this point it is interesting to make a quick sidestep and  note  that
Titov and Beenakker (Ref.~\onlinecite{titov:041401})
showed that for short junctions, at zero temperature and at the Dirac point
the C$\Phi$R for ballistic graphene is formally  identical to the classical result
of  Kulik and Omel'yanchuk\cite{kulik-classical},  which, however,
assumes diffusive normal metal as a weak link.
Looking at  the $I_c-T$ curves in the short junction limit
[ Fig.~\ref{kT_depend:fig}(a)]  one can see that they  are
also qualitatively similar to the corresponding result of Kulik and Omel'yanchuk\cite{kulik-classical}
[c.f. Fig.~7 in Ref.~\onlinecite{CPR-RevModPhys}]. 
The close resemblance of certain properties
the two types of Josephson  junctions therefore seems to extend to the temperature
dependence of the critical current as well.
In the opposite, long junction regime, for $\mu=0$  [shown by red circles in  Fig.~\ref{kT_depend:fig}(c)]
one can observe a short plateau in the current for small temperatures followed by an exponential decay.
Interestingly, a qualitatively very similar result has been obtained
by Gonz\'alez and Perfetto in Ref.~\onlinecite{JPhysC.20.145218}, assuming  tunnel
coupling between the superconductor and the graphene and using a different formalism than ours.
In the case of  finite doping [black squares in  Fig.~\ref{kT_depend:fig}(c)] an exponential
decay of $I_c$ can be seen basically in the whole temperature range.
The $I_c$ in clean SNS junction exhibits the same qualitative dependence on $T$
(Ref.~\onlinecite{CPR-RevModPhys}).

Experimentally,  the $I_c-T$ relation was measured by Du \emph{et al.}
(Ref.~\onlinecite{du:184507}) and by  Ojeda-Aristizabal \emph{et al.}
(Ref.~\onlinecite{ojeda}). Again, quantitative comparison with our results is
not possible because the graphene samples in the experiments were in the diffusive
limit, but the observed dependence of $I_c$ on the temperature was qualitatively similar
to the results shown in  Fig.~\ref{kT_depend:fig}(a).

\begin{figure}[!ht]
\includegraphics[scale=0.5]{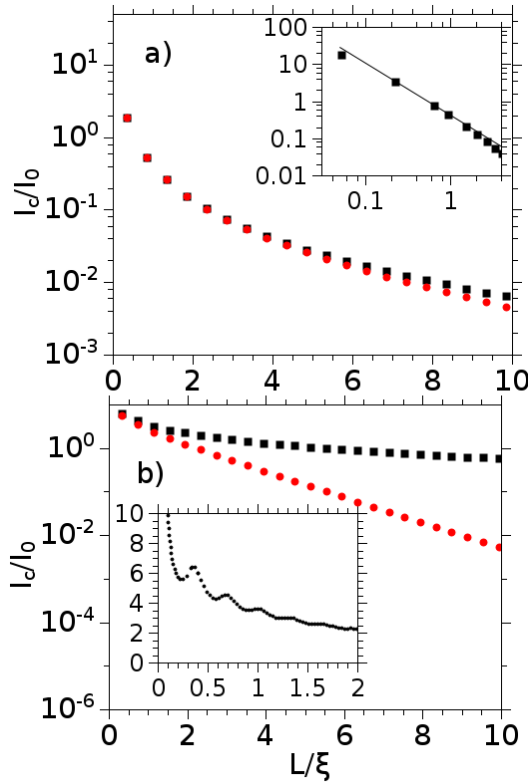}
\caption{(color online)
The critical current $I_c$ 
as a function of the junction length $L/\xi$ in logarithmic scale.
In (a) we used $\mu=0$. Black $\Box$  and red $\circ$ denote the results of
$T/T_c = 0.0$  and $T/T_c = 0.06$ calculations, respectively.
The symbols in the inset of (a) show   $I_c$ for $T/T_c = 0.0$,
$L/\xi < 1$ in double logarithmic plot, along with the fitted linear function
(solid line, see text).  In (b) the chemical potential is $\mu/\Delta_0(0)=10$.
Black $\Box$  and red $\circ$ denote the results of
$T/T_c = 0.0$  and $T/T_c = 0.18$ calculations, respectively. The inset of
(b) shows the $T/T_c = 0.0$ calculations in linear scale for $L/\xi \le 2$.
The width of the junction was $\xi/W=0.05$ in all cases.
\label{fig:L_depend}}
\end{figure}

Finally, we  studied  the length dependence of the critical current
and the results are shown in Fig.~\ref{fig:L_depend}.
At the Dirac point ($\mu=0$) one can observe an exponential decay of $I_c$
for $L/\xi\gg 1$ [main panel of Fig.\ref{fig:L_depend}(a)].
We could not see the $\sim 1/L^2$ dependence predicted in Ref.~\onlinecite{JPhysC.20.145218}.
 For  $L/\xi\lesssim 1$ and $T/T_c=0$, however, we do find
a power-law dependence $I_c\sim L^b $ [see the inset of Fig.\ref{fig:L_depend}(a)] and
fitting the numerical results  we obtained $b=-1.4$. This is remarkably close to
the results of the self-consistent tight-binding
calculations of Black-Schaffer and Doniach (Ref.~\onlinecite{PhysRevB.78.024504})
who found $b=-1.3$.
Considering now the case of doped graphene weak link, we found that for $L/\xi > 1$
the $L$ dependence of $I_c$ can be well fitted  by $I_c/I_0 = I_a e^{- b\, L/\xi}$ with
$b\approx 1$. Exponentially small critical current is also typical for clean SNS junctions
if $L$ is larger than the thermal coherence length (for details see e.g.
Ref.~\onlinecite{CPR-RevModPhys}).
We note that in the case of $L/\xi < 1$ and $T/T_c=0$ one can see oscillations in
the $I_c$ vs $L$ curve [shown in the inset of Fig.\ref{fig:L_depend}(b)]
whose study is left to a future work.

\section{Conclusions}
\label{summary}
In this work we calculated the Josephson current in ballistic SGS structures.
The most important assumption we made is that one can use rigid boundary conditions. i.e. that
the change of the pair potential is step-function-like at the SG interfaces and that it does
not depend on the supercurrent. We developed a  general  and numerically efficient
approach to obtain the current-phase relation for arbitrary length of the junction
as well as for finite doping and temperature.
At low temperatures we have found that the current-phase relation differs
from a simple harmonic dependence.
In the case of  short junctions and  small critical currents the deviation of the current-phase curve
from the sinusoidal form, the skewness,
shows a linear dependence on the critical current, similarly to the observation of a
recent experiment\cite{Exp_Chialvo_graphene_supra:cikk}, though the slope of the curve did
not match the experimental one.  This is likely to  be due to the fact that in the experiment the
graphene sample was quasi-diffusive.
In the long junction limit our results show that in clean SGS junctions the
the current-phase relation transforms from the sinusoidal form at $T\lesssim T_c$ to a curve resembling
saw-tooth  at $T\ll T_c$. In contrast to clean SNS junctions however, our numerics suggests  that the
dependence is not exactly saw-toothed.
We have also calculated the temperature and junction length dependence of the critical current.
We have found similarities to both classical results for SNS junctions
and recent ones obtained for graphene but using a different formalism\cite{JPhysC.20.145218}.
In respect of these numerical calculations further theoretical progress is needed to unravel
the relation  between the SGS and SNS results.

Since the fabrication  of both  ballistic graphene samples  and transparent
SG interfaces have already been demonstrated experimentally\cite{Miao:cikk,du:184507}, we
believe that our theoretical approach may be useful in the future to understand and
analyze  experimental data. In particular, the measurement of the
C$\Phi$R and the length and temperature dependence of $I_c$ in short junctions should be feasible.

\emph{Note added}: During the peer-review process of the manuscript, a relevant preprint has
appeared\cite{black-schaffer} where the authors study the temperature dependence of the Josephson
current using self-consistent tight-binding numerical computations. Their work is thus complementary
to our and help to understand the scope of certain approximations, e.g. the rigid-boundary condition
that we employed.

\noindent \emph{Acknowledgements:} Supported by the Marie Curie ITN project NanoCTM (FP7-PEOPLE-ITN-2008-234970)
and the Hungarian Science Foundation OTKA under the contracts No.~75529 and No.~81492.
A. K. also acknowledges the support of EPSRC.
We acknowledge fruitful discussions with C. W. J. Beenakker, E. Perfetto,  N. Mason,
D. J. Van Harlingen and  C. Chialvo.

\end{document}